\newcommand{\nn}{\nonumber}
\newcommand{\p}[1]{(\ref{#1})}
\newcommand{\pl}{\partial}
\renewcommand{\L}{{L}}
\documentstyle[11pt]{article}
\begin{document}
\thispagestyle{empty}
\begin{flushright} JHU-TIPAC-97012\\ July 1997 \end{flushright}
\bigskip
\bigskip
\begin{center}
{\bf \Large{The tensor Goldstone multiplet \\[3mm]
for partially broken supersymmetry }}
\end{center}
\vskip 1.0truecm

\centerline{\bf Jonathan Bagger} \vskip3mm
\centerline{and} \vskip3mm
\centerline{\bf Alexander Galperin}
\vskip3mm

\centerline{Department of Physics and Astronomy}
\centerline{Johns Hopkins University}
\centerline{Baltimore, MD 21218 USA}
\vskip3mm
\bigskip \nopagebreak

\begin{abstract}
\noindent
We show that the tensor gauge multiplet of $N=1$ supersymmetry can serve
as the Goldstone multiplet for partially broken rigid $N=2$ supersymmetry.
We exploit a remarkable analogy with the Goldstone-Maxwell multiplet of
\cite{Maxwell} to find its nonlinear transformation law and its invariant
Goldstone action.  We demonstrate that the tensor multiplet has two
dualities.  The first transforms it into the chiral Goldstone multiplet;
the other leaves it invariant.
\end{abstract}

\newpage\setcounter{page}1
\begin{flushright}
{\it To the memory of Viktor I. Ogievetsky}
\end{flushright}

\noindent 1.  As with any continuous symmetry, the spontaneous breaking
of supersymmetry implies the existence of a Goldstone field.  For the
partial breaking of extended supersymmetry, this field -- a spin-1/2
Goldstone fermion $\psi_\alpha(x)$ -- belongs to a massless multiplet
of the residual unbroken supersymmetry.

Curiously enough, the choice of Goldstone multiplet is not unique.  For
example, when $N=2$ supersymmetry is spontaneously broken to $N=1$, the
Goldstone field can be a member of a chiral $N=1$ multiplet \cite{BG}.
(This corresponds to the supermembrane of ref.~\cite{Pol}).  In a recent
paper \cite{Maxwell}, we demonstrated that the Goldstone field can also
be contained in the $N=1$ Maxwell spin-(1/2,1) multiplet.  We derived
its nonlinear transformation under the spontaneously broken supersymmetry
and constructed the invariant action.  We found the Goldstone action
to be duality invariant, and its spin-1 part to be
precisely the Born-Infeld action \cite{BI}.

In this letter we will show that there is yet another Goldstone multiplet
for partially broken $N=2$ supersymmetry -- the $N=1$ tensor gauge
multiplet (whose field strength is the $N=1$ linear multiplet)
\cite{linear}.  The superpartners of the Goldstone
fermion are a real scalar field $\ell(x)$ and a real antisymmetric tensor
gauge field $E_{mn}$.  We will see that the $N=1$ description of this
multiplet is almost identical to that of the Maxwell multiplet.  Using
this fact and results of \cite{Maxwell}, we will derive the second,
nonlinear supersymmetry, as well as the invariant action for the tensor
Goldstone multiplet.  We shall see that the tensor multiplet has two
different dualities.  One relates it to the chiral Goldstone multiplet
of \cite{BG}; the other maps it to itself.  We will conclude with a
brief discussion of some unanswered questions, especially those relating
to higher symmetries associated with the tensor Goldstone multiplet.

\vspace{3mm}

\noindent 2. The $N=1$ tensor multiplet is usually described by a real
scalar superfield $\L(x,\theta,\bar\theta)$, constrained by
\begin{equation}
D^2 \L=\bar D^2 \L = 0.
\label{L-constraint}
\end{equation}
Its independent components are the $\theta = 0$ projections of $\L$, $D_\alpha
\L $ and $[D_\alpha, \bar D_{\dot\alpha}]\L$.  They correspond to a scalar
$\ell(x)$, a fermion $\psi_\alpha(x)$ and the field strength of a gauge
tensor $V^m = {1\over 2}\epsilon^{mnkl}\partial_n E_{kl}$.

The solution to \p{L-constraint} is given by
\begin{equation}
\L = D^\alpha \phi_\alpha + \bar D_{\dot\alpha} \bar\phi^{\dot\alpha}\ ,
\label{L-solution}
\end{equation}
where $\phi_\alpha$ is an arbitrary chiral $N=1$ superfield, $\bar
D_{\dot\alpha}\phi_\alpha = 0$.  For our purposes, we find it more instructive
to describe the tensor multiplet in terms of a {\it spinor} superfield
$\psi_\alpha$,
\begin{equation}
\psi_\alpha =  iD_\alpha \L.
\label{psi-from-L}
\end{equation}
Using (\ref{L-constraint}), we see that $\psi_\alpha$ is {\it antichiral}
\cite{BW},
\begin{equation}
D_\beta\psi_\alpha =  0.
\label{psi-antichiral}
\end{equation}
In addition it satisfies the constraint
\begin{equation}
\bar D^2\psi_\alpha - 4i\pl_{\alpha\dot\alpha}\bar\psi^{\dot\alpha}= 0.
\label{irred}
\end{equation}
Equations (\ref{psi-antichiral}) and (\ref{irred}) are equivalent to
(\ref{L-constraint}); they are irreducibility conditions for the
superspin-$1/2$ representation of $N=1$ supersymmetry.

The free action of the tensor multiplet can be written as
\begin{equation}
S_{\rm free}(\psi) = {1\over 4}\int d^4xd^2\theta\, \bar\psi^2 +
{1\over 4}\int d^4xd^2\bar\theta\, \psi^2 .
\label{free_action_linear}
\end{equation}
In this spinorial formulation, the $N=1$ tensor multiplet displays a
remarkable similarity to the $N=1$ Maxwell multiplet.  As is well known,
the latter is described by a {\it chiral} spinor superfield $W_\alpha$,
\begin{equation}
\bar D_{\dot\alpha} W_\alpha = \ 0,
\label{W-chiral}
\end{equation}
which satisfies the reality constraint
\begin{equation}
D^\alpha W_\alpha + \bar D_{\dot\alpha} \bar W^{\dot\alpha} = \ 0.
\label{W-reality}
\end{equation}
This reality constraint (\ref{W-reality}) can be rewritten as an
irreducibility condition,
\begin{equation}
D^2 W_\alpha - 4i\pl_{\alpha\dot\alpha}\bar W^{\dot\alpha}= \ 0.
\label{W-irred}
\end{equation}

The free action of the Maxwell multiplet reads
\begin{equation}
S_{\rm free}(W) = {1\over 4}\int d^4xd^2\theta\, W^2 +
{1\over 4}\int d^4xd^2\bar\theta\, \bar W^2 .
\label{free_action_Maxwell}
\end{equation}
Comparing equations \p{psi-antichiral}, \p{irred}, \p{free_action_linear}
with \p{W-chiral}, \p{W-irred}, \p{free_action_Maxwell}, we see a very
close analogy between the tensor and Maxwell multiplets.  One theory is
related to the other by

\vspace{2mm}
\noindent
(i) switching chirality in $N=1$ superspace:
\begin{equation}
x^m \rightarrow x^m, \ \theta^\alpha \rightarrow \bar\theta^{\dot\alpha},
\ \bar\theta^{\dot\alpha} \rightarrow \theta^\alpha,
\label{switch}
\end{equation}
which maps the chiral $N=1$ superspace $(x^m -2i\theta\sigma^m\bar\theta,
\theta^\alpha)$ to the antichiral superspace $(x^m +2i\theta\sigma^m\bar\theta,
\bar\theta^{\dot\alpha})$; and

\vspace{2mm}
\noindent
(ii) keeping spinor superfield $\psi_\alpha$ inert:
\begin{equation}
\label{inert}
\psi_\alpha(x - 2i\theta\sigma\bar\theta, \theta) \rightarrow
\psi_\alpha(x + 2i\theta\sigma\bar\theta, \bar\theta).
\end{equation}

\vspace{2mm}
\noindent
In what follows we will use this analogy to establish the main features of
the Goldstone tensor multiplet.

\vskip3mm

\noindent 3. To find the broken supersymmetry and invariant action for
the tensor multiplet, we extract the corresponding features of the
Goldstone-Maxwell multiplet from \cite{Maxwell} and then switch chirality.

Using this technique, we find that the second supersymmetry for the
tensor multiplet is given by
\begin{equation}
\delta\psi_\alpha = \eta_\alpha - {1\over 4} D^2\bar X \eta_\alpha
-i\pl_{\alpha\dot\alpha}X\bar\eta^{\dot\alpha},
\label{S_susy}
\end{equation}
which implies
\begin{equation}
\delta \L =  i(\bar\theta\bar\eta - \theta\eta)
		-{i\over 2} D^\alpha \bar X \eta_\alpha
		+{i\over 2} \bar D_{\dot\alpha} X
                \bar\eta^{\dot\alpha}.
\label{S_susy_L}
\end{equation}
Here $X$ is an antichiral $N=1$ superfield, $D_{\alpha} X = 0$,
which satisfies the recursive equation
\begin{equation}
X = {\psi^2 \over 1 - {1\over 4} D^2 \bar X},
\label{X_constraint}
\end{equation}
with the solution
\begin{eqnarray}
X\ &=& \psi^2\ +\ {1\over 2} D^2\left[
{\psi^2\bar\psi^2 \over 1 - {1\over 2}  A\ +\ \sqrt{1-A+{1\over 4}B^2}}
\right] . \\
A &=&{1\over 2}(\bar D^2\psi^2 + D^2\bar\psi^2), \nn\\
B &=& {1\over 2}(\bar D^2\psi^2 - D^2\bar\psi^2).
\label{X_for_W}
\end{eqnarray}
Under the second supersymmetry, $X$ transforms as follows,
\begin{equation}
\delta X = 2\psi\eta.
\label{S_susy_for_X}
\end{equation}
The transformations \p{S_susy} and \p{S_susy_for_X} are consistent with
the contraints imposed on the superfields; they close into the $N=2$
supersymmetry algebra.

Using these results, it is not hard to show that the invariant Goldstone
action is given by
\begin{equation}
S_{\rm tensor} = {1\over 4} \int d^4xd^2\theta\, \bar X + {1\over 4}
\int d^4xd^2\bar\theta\, X.
\label{Xaction}
\end{equation}
This is just the chirality-switched Goldstone-Maxwell action.  Note
that the bosonic part of the action is
\begin{equation}
S_{\rm Bose} = \int d^4x \left[1 - \sqrt{1 - (\pl^m \ell\pl_m \ell - V^mV_m)
- (V^m\pl_m \ell)^2}\ \right].
\label{bose_action}
\end{equation}
If $V_m$ were zero, this would be the action of a 3-brane in five
dimensions.

\vskip3mm
\noindent 4. The tensor multiplet enjoys two different dualities.  At the
level of the bosonic fields, they can be understood as follows.  In four
dimensions, a tensor gauge field is dual to a scalar field and vice versa.
Given a real tensor gauge field $E_{mn}$, one can dualize to a real scalar
field $\tilde\ell$ by relaxing the constraint on its field strength $V^m$,
$\partial_m V^m = 0$, and adding the Lagrange multiplier
\begin{equation}
\int d^4x \, V^m \partial_m \tilde \ell.
\end{equation}
Alternatively, one can transform from a real scalar $\ell$, with ``field
strength'' $\ell_m = \partial_m \ell$, to a tensor gauge field $\tilde{E}_{mn}$
by relaxing the constraint $\partial_{[m}\ell_{n]} = 0$, and adding the
Lagrange multiplier
\begin{equation}
\int d^4x \, \epsilon^{mnkl} \tilde{E}_{kl}\partial_{[m}\ell_{n]}.
\end{equation}

Using these techniques, one can dualize the field strength $V_m$ to a scalar
field in the bosonic part of the Goldstone action \p{bose_action}.  This leads
to a dual action with two physical scalar fields.  Alternatively, one can
dualize the scalar field $\ell$ {\it and} the tensor field strength $V_m$.
The dual action has the same field content -- a scalar and a gauge tensor --
as the original action.  One can check that the first duality transforms
\p{bose_action} into the action for a 3-brane in six dimensions, while the
second leaves it invariant.

In superspace, the second duality can be demonstrated using the spinorial
representation for the tensor multiplet, and relaxing the irreducibility
constraint \p{irred} while keeping $\psi_\alpha$ antichiral.  This is
done by introducing the Lagrange multiplier
\begin{equation}
i\int d^4x d^2\theta
\, \widetilde \psi^\alpha \psi_\alpha - i\int d^4x d^2\bar\theta
\, \widetilde{\overline\psi}_{\dot\alpha} \overline\psi^{\dot\alpha},
\label{Lagr_psi}
\end{equation}
where $\widetilde \psi^\alpha$ is an antichiral superfield, subject
to the irreducibility constraint \p{irred},
\begin{equation}
\bar D^2\widetilde\psi_\alpha - 4i\pl_{\alpha\dot\alpha}
\widetilde{\overline\psi}{}^{\dot\alpha} =  0.
\label{tilde_irred}
\end{equation}
The relaxed action \p{Xaction}, \p{Lagr_psi} has exactly the same form
as the relaxed action for the Goldstone-Maxwell multiplet \cite{Maxwell}
(after switching chirality).  Therefore, as in \cite{Maxwell}, varying
with respect to $\psi$ and substituting back produces the dual action
\begin{equation}
S(\widetilde\psi) = {1\over 4} \int d^4xd^2\theta
\, \widetilde{\overline X} +\ {1\over 4} \int
d^4xd^2\bar\theta\, \widetilde X,
\label{dual_action}
\end{equation}
where
\begin{equation}
\widetilde X =
{\widetilde\psi^2 \over 1 - {1\over 4} D^2 \widetilde{\overline X}}.
\label{tilde_X_constraint} \end{equation}
The dual action has exactly the same form as the original action \p{Xaction}.

The superspace form of the first duality maps the $N=1$ tensor multiplet
into its chiral counterpart.  The duality is implemented by rewriting the
action \p{Xaction} in terms of the real superfield $\L$, and then relaxing
the constraint \p{L-constraint} by adding the Lagrange multiplier
\begin{equation}
\int d^4xd^4\theta\, \L(\phi + \bar\phi),
\label{Lagr_phi}
\end{equation}
where $\phi$ is a chiral superfield, $\bar D_{\dot\alpha}\phi = 0$.  If
one varies the relaxed action with respect to $\L$ and substitutes back,
one obtains the full nonlinear action for the chiral Goldstone multiplet
\cite{RT},
\begin{equation}
S_{\rm dual}(\phi) \ =\  \int d^4xd^4\theta\ {\cal L}(\phi,\bar\phi),
\label{S_phi}
\end{equation}
where
\begin{equation}
{\cal L}(\phi,\bar\phi) \ =\  \phi\bar\phi + {1\over 8}
\,D^\alpha\phi D_\alpha\phi\bar
D_{\dot\alpha}\bar\phi\bar D^{\dot\alpha}\bar\phi\,f
\end{equation}
and
\begin{eqnarray}
f^{-1} &=& 1 + {A\over 2} + \sqrt{1 + A + B}, \nonumber\\
A &=& -4 \pl_m \phi \pl^m \bar\phi - {1\over4} D^2\phi {\bar D}^2 \bar\phi,
\nonumber\\
B &=& 4 (\pl_m \phi \pl^m \bar\phi)^2 - 4 (\pl_m\phi)^2 (\pl^n\bar\phi)^2.
\end{eqnarray}
The action is invariant under the full nonlinear second supersymmetry,
\begin{equation}
\delta\phi\ =\ -2i \theta\eta - {i\over4} \eta^\alpha {\bar D}^2 D_\alpha
{\cal L}.
\end{equation}
This transformation closes into the off-shell $N=2$ supersymmetry algebra.
Changing variables as follows,
\begin{equation}
\phi\ =\ \varphi + {1\over 16} \bar D^2(\bar\varphi
D^\alpha\varphi D_\alpha\varphi) + O(\varphi^5)
\label{change_phi}
\end{equation}
one can show that the leading terms of the action \p{S_phi} are precisely
those of the chiral Goldstone action derived in \cite{BG}.

It is interesting to note that the superfield ${\cal L}$ plays the role of
an $N=1$ Maxwell prepotential.  Indeed, if we define the field strength,
\begin{equation}
{\cal W}_\alpha\ =\ -{i\over 8} {\bar D}^2 D_\alpha {\cal L},
\end{equation}
we find that $\phi$ and ${\cal W}$ obey the following $N=2$
transformation laws,
\begin{eqnarray}
\delta\phi &=& - 2 i \theta\eta + 2 {\cal W} \eta, \nonumber\\
\delta {\cal W}_\alpha &=& -{1\over 4} {\bar D}^2 \bar\phi\, \eta_\alpha
- i \partial_{\alpha\dot\alpha}\phi\, \bar\eta^{\dot\alpha}.
\label{S_susy_for_W}
\end{eqnarray}
We see that the Goldstone field $\phi$ and its $N=1$ superfield
Lagrangian are elements of an $N=2$ vector multiplet.  Together
they form a {\it linear} representation of $N=2$ supersymmetry --
up to a field-independent shift.

In fact, the same holds true for the other Goldstone multiplets
as well.  Comparing \p{S_susy_for_W} with eqs.~(27) and (28) of
ref.~\cite{Maxwell}, and with eqs.~\p{S_susy} and \p{S_susy_for_X}
above, we see that the nonlinear transformation laws take a
similar form.  The vector multiplet Goldstone field, $W_\alpha$,
and its superfield Lagrangian, $X$, combine to form an
vector $N=2$ multiplet.  Their transformations are identical
to \p{S_susy_for_W} (ignoring the shifts, and replacing
$\phi$ by $X$ and ${\cal W}_\alpha$ by $W_\alpha$).  The tensor
multiplet Goldstone superfield, $\psi_\alpha$, and its Lagrangian,
$X$, are the elements of an $N=2$ tensor multiplet.

\vskip3mm
\noindent 5. The analogy between the $N=1$ tensor gauge and Maxwell multiplets
is closely related to an early work of Ogievetsky and Polubarinov \cite{OP},
in which the authors introduced an antisymmetric tensor gauge field (the
``notoph''), which is complementary to the photon.  The notoph also has
three degrees of freedom, but propagates with helicity zero on shell.
Our chirality flipping procedure takes the $N=1$ Maxwell multiplet into
the $N=1$ tensor gauge multiplet, which is the supersymmetric
generalization of the notoph.

By construction, the Goldstone action \p{Xaction} is invariant under
$N = 2$ supersymmetry.  It turns out that this action is also invariant
under a full five-dimensional Poincar\'e supersymmetry.  The scalar
$\ell(x)$ is the Goldstone boson associated with the momentum in the fifth
dimension; from a four-dimensional point of view, it is the Goldstone
boson for a real central charge (as can be seen from \p{S_susy_L}).
Furthermore, the gradient $\partial_m \ell(x)$ can be shown to parametrize
the Lorentz group coset $SO(1,4) / SO(1,3)$.

The fact that $\ell(x)$ and $\psi_\alpha(x)$ are both Goldstone fields
leads one to speculate that the antisymmetric tensor gauge field might
itself be a Goldstone field.  If so, it would be interesting to
understand its role from the algebraic and $p$-brane points of view.

This work completes a series of three papers in which we showed
that the Goldstone fermion from the spontaneous breaking of $N=2$
supersymmetry can be an element of an $N=1$ chiral, vector or tensor
multiplet.  However, these papers do not answer the most intriguing
question of all:  Why are there three such multiplets, when only one
would do?

J.B. would like to thank Arkady Tseytlin for a preliminary version of
ref.~\cite{RT}.  This work was
supported by the U.S. National Science Foundation, grant NSF-PHY-9404057.

\end{document}